\documentclass[10pt]{article}
\usepackage{mathbbold}
\usepackage{multirow}

\usepackage{amssymb}
\usepackage{graphicx}
\usepackage{threeparttable}
\usepackage{tabularx}
\usepackage{array}
\usepackage{dcolumn}
\usepackage{bm}

\usepackage[labelfont=bf,labelsep=period,justification=raggedright]{caption}

\makeatletter
\renewcommand{\@biblabel}[1]{\quad#1.}
\makeatother

\date{\today}

\newcommand{\PreserveBackslash}[1]{\let\temp=\\#1\let\\=\temp}
\newcolumntype{C}[1]{>{\PreserveBackslash\centering}p{#1}}
\newcolumntype{R}[1]{>{\PreserveBackslash\raggedleft}p{#1}}
\newcolumntype{L}[1]{>{\PreserveBackslash\raggedright}p{#1}}

\begin{document}

\begin{flushleft}
{\Large \textbf{Information Filtering on Coupled Social Networks}
}
\\Da-Cheng Nie$^{1}$,
Zi-Ke Zhang$^{2,3,4,\ast}$,
Jun-lin Zhou$^{1}$,
Yan Fu$^{1}$,
Kui Zhang$^{2}$
\\
\bf{1} Web Sciences Center, School of Computer Science \& Engineering, University of Electronic Science and Technology of China, Chengdu 610054, People's Republic of China
\\
\bf{2} College of Communication Engineering, Chongqing University, Chongqing 400044, China
\\
\bf{3} Institute of Information Economy, Hangzhou Normal University, Hangzhou 311121, People's Republic of China
\\
\bf{4} Alibaba Research Center for Complexity Sciences, Hangzhou Normal University, Hangzhou 311121, People's Republic of China
\\
$\ast$ Corresponding Author: zhangzike@gmail.com
\end{flushleft}


\section*{Abstract}

In this paper, based on the coupled social networks (CSN), we propose a hybrid algorithm to nonlinearly integrate both social and behavior information of online users. Filtering algorithm based on the coupled social networks, which considers the effects of both social influence and personalized preference. Experimental results on two real datasets, \emph{Epinions} and \emph{Friendfeed}, show that hybrid pattern can not only provide more accurate recommendations, but also can enlarge the recommendation coverage while adopting global metric. Further empirical analyses demonstrate that the mutual reinforcement and rich-club phenomenon can also be found in coupled social networks where the identical individuals occupy the core position of the online system. This work may shed some light on the in-depth understanding structure and function of coupled social networks.

\section*{Introduction}




In the past two decades, the rapid development of Internet has provided an unlimited source for us to search and find what we need \cite{faloutsos1999}. For instance, we now can enjoy plenty of TV channels as well as countless programs, while only few choice is available twenty years ago. Moreover, the \emph{Internet} not only offers various games, but also becomes a versatile tool to change the lifeway that we have kept constantly over centuries. For example, online shopping has become more and more popular due to the exponential growth of e-commerce services (e.g. \emph{Amazon.com}, \emph{Ebay.com}, \emph{Taobao.com}, etc), which allow us to choose, compare and purchase goods with single clicks. In addition, there is a vast class of novel job portions arising with the emergence of web related applications, such as \emph{SOHO} workers (working at home but communicating via Internet). However, everything has two sides. Although Internet has changed the world a lot and much improved our lifespan to effectively and efficiently contact with others, it also brings many side effects, some of which are becoming critically important and even  disruptive to our day-to-day routines. One of the most significant dilemmas is the well-known \emph{Information Overload} problem. Take the aforementioned TV programs for example. In despite of the fact that we indeed have more items to choose than ever before, it is simultaneously surprising to see that we are even more difficult to find a proper  program satisfying us. That is to say, we are facing too many choices to be able to compare each other and make the appropriate decision.

Recently, researchers from various disciplines, including computer science, social science, physics, etc., have devoted much effort to helping users avoid being drowned into the \emph{Information Ocean} \cite{froomkin1995}. Among numerous applications, the most successful milestone is the emergence of \emph{Search Engine} (SE) \cite{brin1998anatomy}, which can help users locate targets by filtering irrelevant objects with designed keywords, hence soon be widely applied over the Internet. Despite its great success in information filtering, the SE technology also has some apparent drawbacks, which interfere its further application in modern human society. On one hand, SE does not consider the personalization of each user, and return exactly the same results for every query with same keywords, regardless of whatever they have searched before \cite{morita1994}. On the other hand, we need to know priori profiles of targets which, however, normally are not very clear for us when the searching is being performed. In addition, some potential intentions are not very easily to be described and expressed by simple words or sentences for users, hence additionally increase the difficulty of predicting their underlying preferences. Moreover, SE can only work when users proactive submit their queries \cite{lee2008find}, thus, it lacks the power of actively providing results based on users' searching histories and personalized preferences.

As a consequence, \emph{Recommender Systems} (RS), focusing on mining users' potential options, is considered as a promising candidate to address the excessive sources problem in the information era \cite{resnick1997, burke2002, Herlocker2004, Adomavicius2005, lv2012}. RS has achieved a great success in the past a few years because it can significantly help users find relevant yet interesting items for them. A recommender system is able to automatically provide personalized recommendations based on the historical records of users' activities. These activities are usually represented by the connections in a user-object bipartite graph \cite{huang2004,Zhou2007}. The majority of relevant works in this area can be generally classified into six representative fields: i) Collaborative Filtering (CF) \cite{breese1998empirical, sarwar2001item}; ii) Content Based Algorithms (CB) \cite{pazzani2007content}; iii) Probability Based Models \cite{kumar1998recommendation, krestel2009latent}; iv) Dimension Reduced Approaches \cite{sarwar2000application}; v) Network Based Inference (NB); \cite{Zhou2007, Zhang2007}; vi) Hybrid Algorithms \cite{burke2002hybrid, Zhou2010}. CF tends to recommend to users with objects that people with similar tastes and preferences liked in the past. There are two categories respectively considering
user-based \cite{resnick1994grouplens} and object-based \cite{sarwar2001item, linden2003amazon} factors, which should be alternatively applied in different online systems according to their own corresponding properties. For instance, \emph{Amazon.com} is a well-known book service in which the number of books is more stable than the rapid growth of readers, and thus object-based algorithms could achieve more reliable recommendation results \cite{linden2003amazon}. Comparatively, \emph{Del.icio.us}\footnote{http://www.delicious.com/} is a typical user-driven social bookmarking platform \cite{zhang2010hypergraph}, hence user-based algorithm is more suitable and effective \cite{Zhang2010EPL}. Content based methods mainly use text mining techniques to automatically extract out meaningful content and then provide recommendations. Both probability and dimension reduced approaches require much more computational time to obtain the latent variables or vectors \cite{zhang2011tag}. By contrast, network based models, making use of physical dynamics (e.g. random walk \cite{wang2001efficient, leicht2006vertex, lu2011pre}, heat conduction \cite{Zhang2007,liujg2011pre,liujg2012pre}), try to apply node diffusion process \cite{sun2009information} to measure the likelihood of given pair of user and object to be connected. Such methods would be adjusted to consider the effects of those small-degree (saying \emph{cold}) objects \cite{qiu2011item, chen2012promotional} and are especially efficient for recommendation for sparse data sets \cite{Zhou2009}. Hybrid algorithms, which normally do not intend to design new methods, but instead introduce one or more tunable parameters to integrate different models \cite{Zhang2010, Zhou2010}.

Recently, \emph{Social Networks} (SN) \cite{freeman1979centrality} has become a powerful tool to characterize various online social services emerging with various Web 2.0 applications \cite{fu2008empirical} in evolutionary games \cite{nowak2006five, szabo2007evolutionary}, community detection \cite{fortunato2010community} and medical science \cite{kamel2007emerging}, etc. A great many of websites have attracted millions of users daily active online. For example, \emph{Twitter} has more than 1.7 $\times$ 10$^8$ users over the world. \emph{Facebook} has reported more than 900 million users registered with two years. \emph{Sina Weibo}, the largest microblogging service in China, has been involved by almost 1/10 population of China. Therefore, SN provides rich yet meaningful social relations to weigh social similarities among users, hence is expected to be a very useful ingredient to generate more accurate, instructive and explainable recommendation results \cite{kautz1997referral}.

\begin{figure}[htb]
   \begin{center}
      \center  \includegraphics[width=8.5cm]{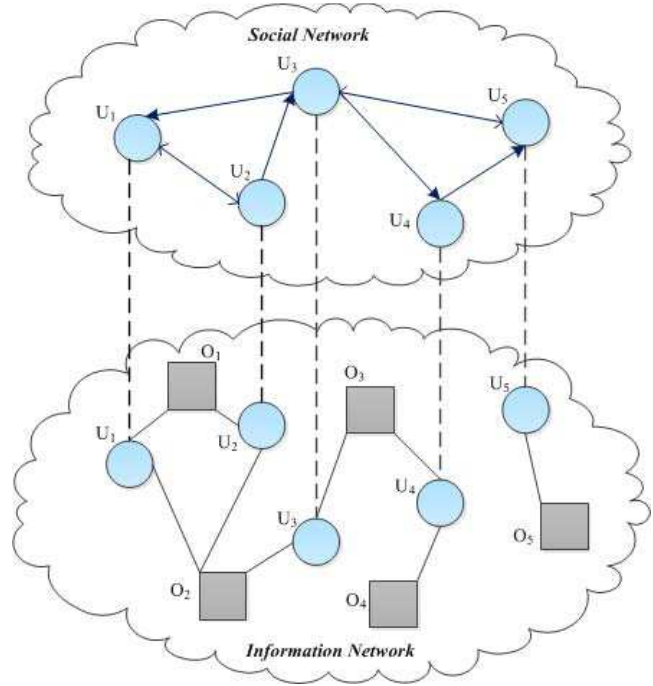}
       \caption{(Color online) Illustration of a coupled social network with five users and five items, where circles denote users and squares represent obejcts. (upper layer) social network consists of five users; (lower layer) the information network consists of five objects and five users, while user nodes are the same in the social network.}\label{fig_illu}
    \end{center}
\end{figure}

Coupled networks (CN), also known as interdependent networks, normally contain a joint two-layer network, such as electricity and Internet networks \cite{buldyrev2010catastrophic}, airport and railway networks \cite{givoni2006airline}. There is a kind of coupled nodes, such as cities in the two aforementioned networks, which play the interconnection and maintenance roles between these two-layer networks. Consequently, those nodes are critically important for the robustness of whole networks \cite{gao2011networks}. Coupled social networks (CSN), similar with the interdependent networks, also contain such coupling nodes (saying users), who both make friends in the layer of social networks and collect favorites in the layer of information networks. Therefore, those users are especially vital for maintaining the structure, connectivity and robustness of social and information networks. Fig. \ref{fig_illu} shows an illustration of a simple CSN with five users and five objects. It is can be seen that the value of similarity between user $U_4$ and user $U_5$ is zero since they do not collect the same object in the information network, which would be considered as there is no relation between them in traditional complex network theory \cite{lu2009similarity}. However, $U_4$ and $U_5$ are friends and may have frequent contacts in the social network, thus they should have many common interests in making acquittance of congenial friends or perform other social activities. Therefore, a reasonable consideration of the similarity of those two nodes should improve the consequent recommendation performance. Massa and Avesani \cite{Massa2007} proposed a social propagation method based on users' distance from a fixed propagation horizon and increased the recommendation coverage while preserving the quality of closeness. There are also many works that introduced social trust and distrust relations to recommender systems \cite{Guha2004,Abdul-Rahman2000}. In \cite{Jsang1998}, the propagation approach was used to combine pairs of trust and distrust. In \cite{Bhuiyan2010} the author discussed the definition of trust, and their results demonstrated the positive relationship between trust and interest similarity in online social networks. \cite{Crandall2008} proposed a feedback effect between similarity and social influence in online communities. Esslimani \emph{et al.} \cite{esslimani2009social} proposed a new information network based collaborative filtering, exploited navigational patterns and transitive links to model users, analyzed behavior similarities, and eventually explored missing links. As we can see, many relationships can constitute a social network such as trust, friendship, community, organizational structure, etc. And some relations are directed, like trust and follower-followee, and others are undirected such as friendship. By utilizing those social relations, we can obtain the strength of social relationship between users, and we can use this weighted social relationship to generate more accurate, explainable and acceptable recommendations when it lacks user behavioral information or their profiles.

With the same motivation, we proposed an algorithm based on CSN by considering the similarities both from social and information networks, and provide recommendation in the classical CF framework. Numerical experiments on two benchmark data sets, \emph{Epinions} and \emph{Friendfeed}, demonstrate that our method can give higher accurate recommendations than previous methods. In addition, extensive analyses show that the RWR-based social similarity can not only enhance the connection between small-degree and large-degree user pairs, but also can reveal the large-distance user pairs which cannot reveled by other direct metrics. As a consequence, a wider range of similar users, which  cannot be discovered solely from information network, could be made of use to generate more reliable yet more precise recommendations.


\section{Methods}\label{Method}

In this section, we start by introducing the approaches of respectively evaluating the social influence and personalized preference between two users. Then, we shill integrate them to measure the final similarity of each pair of users, and apply them in. Generally, a recommender
system consists of two sets, respectively of users $U=\{U_1,U_2,\ldots,U_n\}$,
and items $I=\{I_1,I_2,\ldots,I_m\}$. Denote $R_{m\times n}$ as the adjacent matrix of the user-item
bipartite network, 
of which each element $R_{ij}=1$ if user $U_i$ has collected item $I_j$, and $R_{ij}=0$ otherwise. Analogously, $T_{m\times m}$ is an asymmetric matrix, denoting the directed social network, where $T_{ij}=1$ if the user $U_i$ has linked to user $U_j$, and $T_{ij}=0$ otherwise.

\subsection{Social Influence}
We firstly use the Random Walk with Restart (RWR) \cite{Tong2006,shang2009relevance,yu2014reverse} method to evaluate the social influence of directed networks. Consider a random walker starting at node $i$. At each step, it can move to $i's$ nearest neighbors via directed links with probability $c\in [0,1]$ or returns to node $i$ with probability $1-c$. And the final probability of each node at the stationary state will be considered as their respective peer-to-peer influence with node $i$. 
Denote $T$ as the transition matrix of the directed network, where $T_{ij}=1/k_i$ ($k_i$ is the out-degree of node $i$ if node $i$ and $j$ are linked). So, the final probability of $i$'s influence to others can be defined in a vector manner, $s_{i}^{RWR}$, as

\begin{equation}
\label{eq1}
  \overrightarrow{s_{i}^{RWR}}=(1-c)(1-cT)^{-1}\overrightarrow{e_i} ,
\end{equation}
where $\overrightarrow{e_i}$ is a unit vector with dimension $m\times 1$, and $m$ is the number of users. Besides the RWR metrics, we also emply two typical local methods: \emph{LIN} and \emph{LOUT} to evaluate the social influence, using the adjusted \emph{Jaccad} method, namely \emph{Tanimoto} coefficient \cite{Anderberg1973, Mild2002}, to compute the social influence between two users. They are defined as:\\
\emph{\textbf{LIN}}:

\begin{equation}
\label{eq2}
s_{ij}^{LIN}=\frac{\sum_{k=1}^{m}T_{ki}T_{kj}}
{\sqrt{\sum_{k=1}^{m}T_{ki}^2}+\sqrt{\sum_{k=1}^{m}T_{kj}^2}-\sum_{k=1}^{m}T_{ki}T_{kj}} ,
\end{equation}
\emph{\textbf{LOUT}}:

\begin{equation}
\label{eq3}
s_{ij}^{LOUT}=\frac{\sum_{k=1}^{m}T_{ik}T_{jk}}
{\sqrt{\sum_{k=1}^{m}T_{ik}^2}+\sqrt{\sum_{k=1}^{m}T_{jk}^2}-\sum_{k=1}^{m}T_{ik}T_{jk}} ,
\end{equation}

Then these metrics (Eq. (\ref{eq1}) - Eq. (\ref{eq3})) will be used to quantify how one user influences others. It can be seen that both $s_{ij}^{LIN}$ and $s_{ij}^{LOUT}$ only consider the local information. That is to say, only the common linked nodes of users $i$ and $j$ are taken into account. Comparatively,
$\overrightarrow{s_{ij}^{RWR}}$, from the perspective of dynamic influence flow, considers both the local and global structure of directed networks. Therefore, it is expected to be a promising index to characterize the social influence, hence may provide better recommendation performance.

\subsection{Personalized Preference}

There are many methods to compute the common preference between users or items in recommender systems, and the cosine metric \cite{Kleinberg1999} is one of the most frequently used one \cite{ziegler2013recommender,liu2014discriminant}. It reads

\begin{equation}
\label{eq4}
p_{ij}=\frac{\sum_{k=1}^{n}R_{ik}R_{jk}}
{\sqrt{\sum_{k=1}^{n}R_{ik}^{2}}\sqrt{\sum_{k=1}^{n}R_{jk}^{2}}},
\end{equation}
where $p_{ij}$ is the examined common preference between nodes $i$ and $j$.

\subsection{Hybrid Algorithm}

To fully make use of the effect of both influence and preference of users, we then adopt a nonlinear hybrid method to integrate them. The final similarity between users $i$ and $j$, $S_{ij}$, is denoted as

\begin{equation}
\label{eq5}
S_{ij}=p_{ij}^\alpha \ast s_{ij}^\beta.
\end{equation}


\section{Data \& Metrics}
\subsection{Data set}
In this paper, we use two data sets (datasets are free to download as \textbf{Supporting Information}), \emph{Epinions.com} \cite{Massa2006} and \emph{Friendfeed.com} \cite{Celli2010}, to evaluate the effect of the algorithm. In $Epinions$, it not only allows users to rate items but also permits them to make social connections with others. $Friendfeed$ is a microblogging service built in 2007 and acquired by $Facebook$ in 2009. To alleviate the sparse problem \cite{zeng2013information}, we purify the two data sets by make sure that each user has at least one out-link and 26 in-links (2 for $Friendfeed$ ) in the social network, and each user at least collects 7 items (8 items for the $Friendfeed$ data set) that each item is collected at least 7 times (8 times for $Friendfeed$). Finally, we obtained a purified data set with 4,066 users, 7,649 items, 217,071 social links and 154,122 bipartite links for $Epinions$, and 4,188 users, 5,700items, 386,804 social links and 96,942 bipartite links for $Friendfeed$. Table \ref{tbl:data} shows the basic statistics for two representative data sets).

\subsection{Metrics}
 Every data set is randomly divided into two parts: the training set which is consisted of 90\% entries and the remainings constitute the testing set. For a general recommendation process, the training set is treated as known information to run algorithms and generate corresponding recommendations, while no information in testing set is allowed to use when making recommendations. In addition,n we use four metrics to evaluate in order to give comprehensive understanding of the methods' performance, we consequently employ four different metrics that characterize recommendation performance:

\begin{enumerate}

\item \emph{Precision} \cite{Herlocker2004} .--
\emph{Precision} represents the probability to what extent a selected item is relevant in a given recommendation list, defined as:

\begin{equation}
  \emph{P}_{i}= \frac{N_{rs}^{i}}{L} ,
\end{equation}
where $L$ represents the length of recommendation's list, $N_{rs}^i$ is the number of truly recovered
items for user $i$. We can obtain the precision of whole recommender system by averaging over all individual¡¯s precisions,

\begin{equation}
  \emph{P}= \frac{1}{m}\sum_{i=1}^{m}P_i ,
\end{equation}
where $m$ represents the number of users. Obviously, a higher precision means the more accurate the algorithm is.

\item \emph{Recall} \cite{Herlocker2004} .---
\emph{Recall} represents the probability that a relevant item will be picked from testing set, defined as:

\begin{equation}
  \emph{R}_i=\frac{N_{rs}^i}{N_{p}^i} ,
\end{equation}
where $N_{p}^i$ is the number of items collected by user $i$ in the testing set, and $N_{r}^i$ is the
number of recovered items of user $i$. We then obtain the overall recall of whole recommender system by averaging over all individuals,

\begin{equation}
  \emph{R}= \frac{1}{m}\sum_{i=1}^{m}R_i.
\end{equation}
A higher recall means the more accurate the algorithm is.

\item \emph{F-measure}  \cite{Herlocker2004} ---
The \emph{F-measure} metric is a widely used metric to alleviate the sensitivity of solely usage of precision or recall, defined as,

\begin{equation}
 \emph{F}_i = \frac{2P_{i}  R_{i}}{P_{i} + R_{i}}.
\end{equation}
Anomalously,  we can obtain the \emph{F-measure} of whole system by averaging over all individuals,

\begin{equation}
  \emph{F}= \frac{1}{m}\sum_{i=1}^{m}F_i.
\end{equation}


%

\item \emph{AUC} \cite{Hanely1982} ---
Different from the above three metrics, \emph{AUC} evaluates the likelihood of all items instead of the \emph{TOP} $L$ recommendation. It can be approached with a sampling method
\begin{equation}
  \emph{AUC}=\frac{n'+0.5n''}{n},
\end{equation}
where $n$ is the number of independent sampling, $n'$ is the number of that the predicted score of target item is higher than the score of the randomly selected item, and $n''$ is the times of the target and random items having the same score.  If all the scores are generated from an independent and identical distribution, the \emph{AUC} should be 0.5. Therefore, the value of the \emph{AUC} exceeds 0.5 indicates how much the algorithm performs better than a random prediction.

\end{enumerate}

\section{Results \& Analysis}

\subsection{Experimental Results }

\begin{figure}[htb]
     \begin{center}
       \center  \includegraphics[width=18cm]{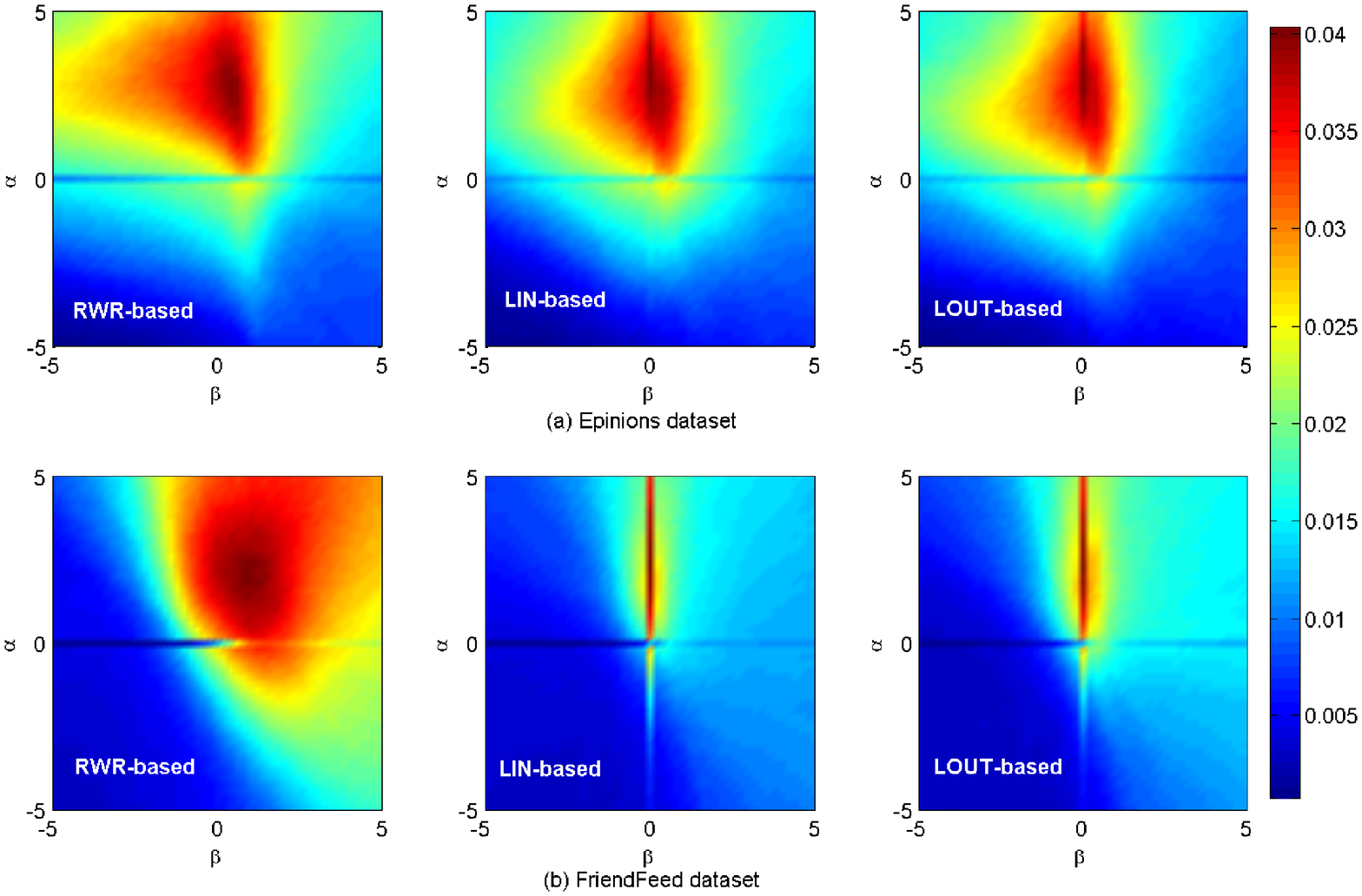}
        \caption{(Color online) \emph{Precision} results on $Epinions$ and $FriendFeed$ data sets. The length of recommendation list $L$ is set as 10.}
    \label{fig_precision}
    \end{center}
\end{figure}

\begin{figure}[htb]
    \begin{center}
       \center \includegraphics[width=18cm]{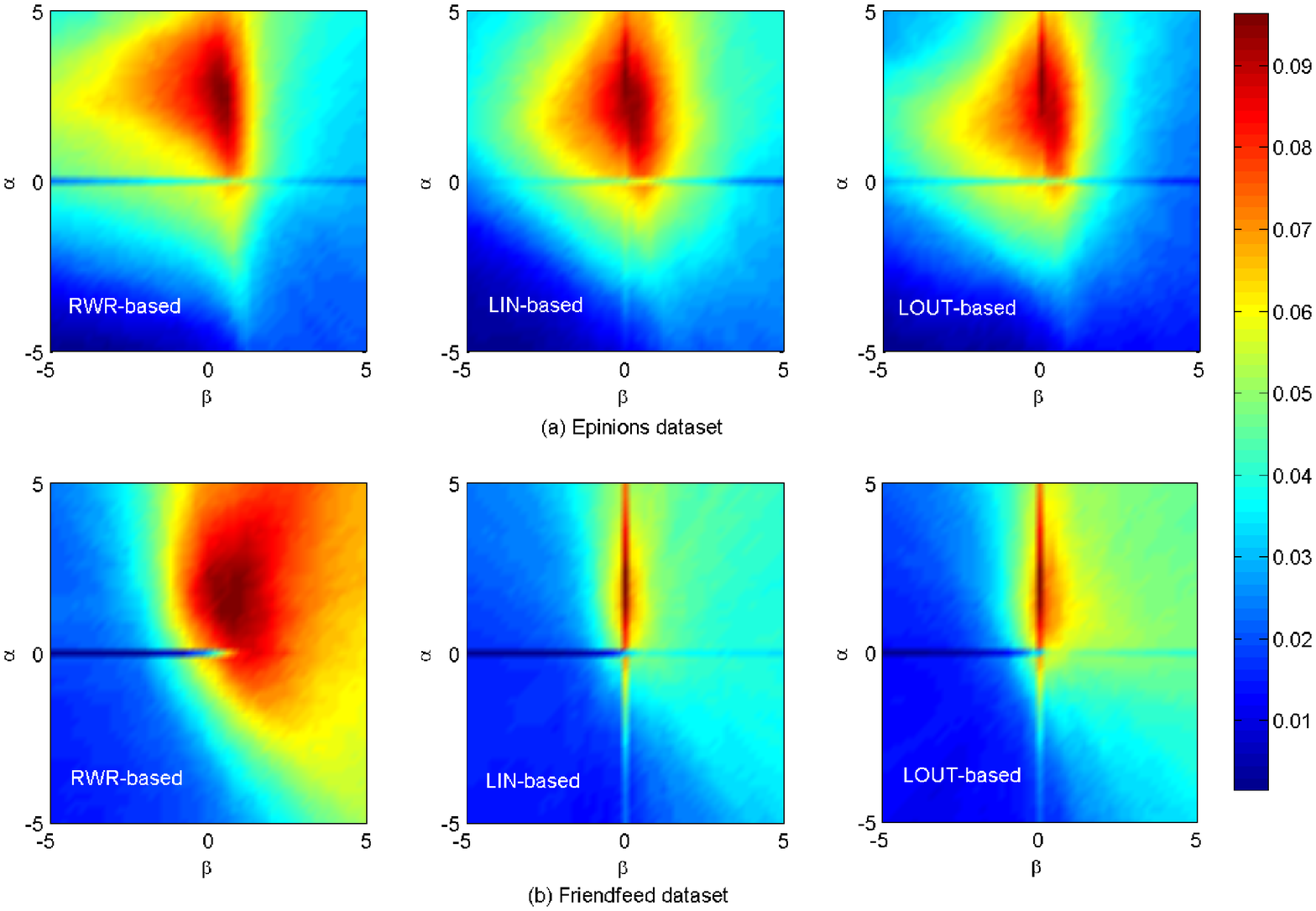}
        \caption{(Color online) \emph{Recall} results on $Epinions$ and $FriendFeed$ data sets. The length of recommendation list $L$ is set as 10.}
    \label{fig_recall}
    \end{center}
\end{figure}

\begin{figure}[htb]
    \begin{center}
       \center \includegraphics[width=18cm]{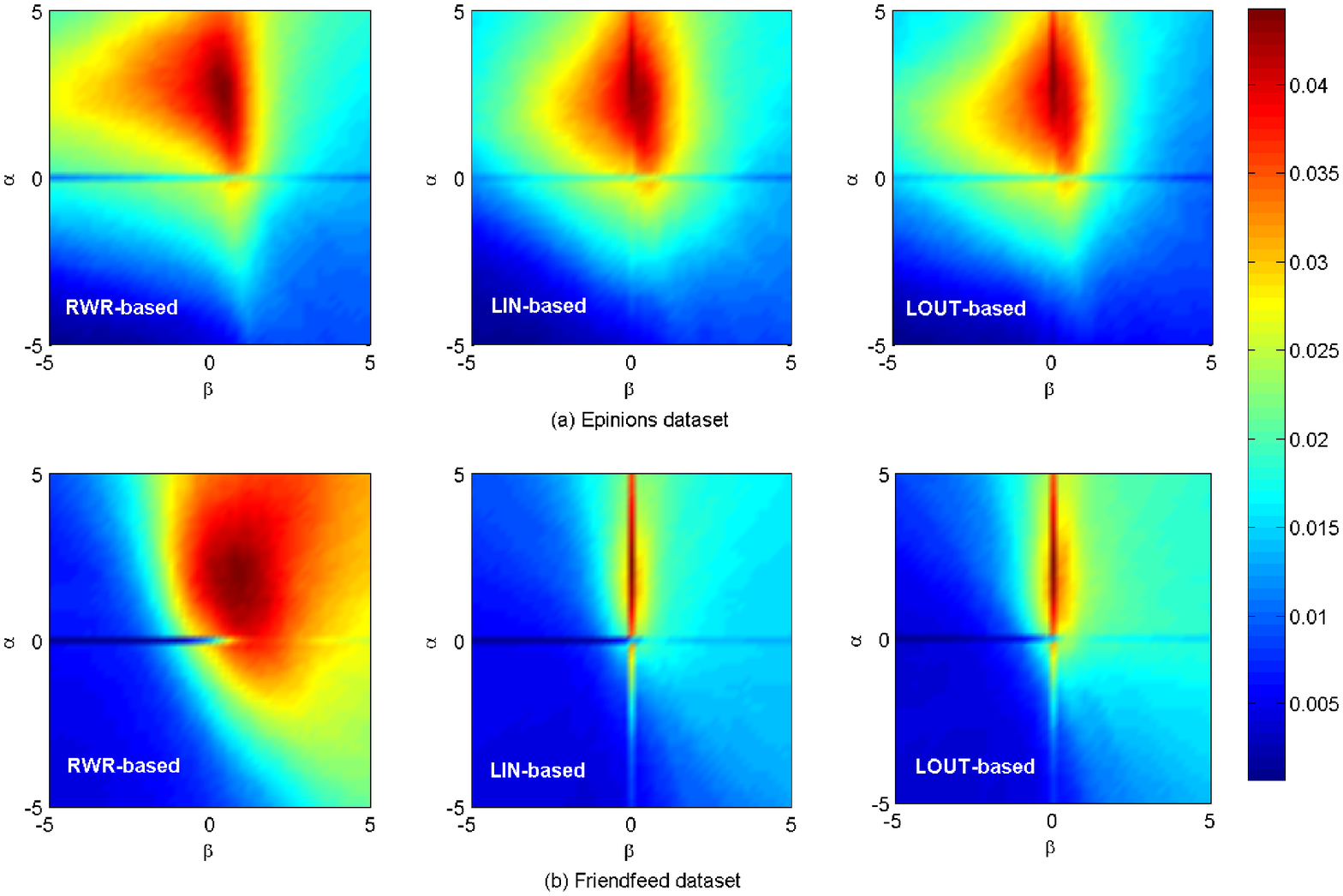}
        \caption{(Color online) \emph{F-measure} results on $Epinions$ and $FriendFeed$ data sets. The length of recommendation list $L$ is set as  10.}
    \label{fig_fmeasure}
    \end{center}
\end{figure}

Fig. \ref{fig_precision} - Fig. \ref{fig_fmeasure} show the algorithm results on $Epinions$ and $Friendfeed$ data sets. It can be seen that, for a given length of recommendation list $L$, the precision, recall, F-measure and AUC obtain the optimal accuracy for the same parameters for both the LIN-based and LOUT-based method (see also Table 2), which indicates that the local information of both in-flow and out-flow have the similar impact in information filtering. Comparatively, for a moderately small length of recommendation list $L$ = 10, the precision, recall and F-measure values of RWR-based method reach their maximum value  0.0526, 0.0717 and 0.0512 for ($\alpha$, $\beta$) = (2.8, 0.4), respectively. And the corresponding results are 0.0503, 0.0683 and 0.0489 for ($\alpha$, $\beta$) = (3, 0) for LIN-based or LOUT-based in $Epinions$ data set. For $Friendfeed$, those metrics under RWR-based method have reached 0.0425, 0.1006 and 0.0469 for parameter set ($\alpha$, $\beta$) = (2, 0.8), (1.4, 0.8) and (2, 0.8), respectively. For LIN-based or LOUT-based methods, when ($\alpha$, $\beta$) = (2.4, 0), such metrics obtain their maximum value 0.0403, 0.0963 and 0.0443. Similar results can also be found for $L=20$ and $L=50$ (see Table 2).

\begin{figure}
    \begin{center}
       \center \includegraphics[width=18cm]{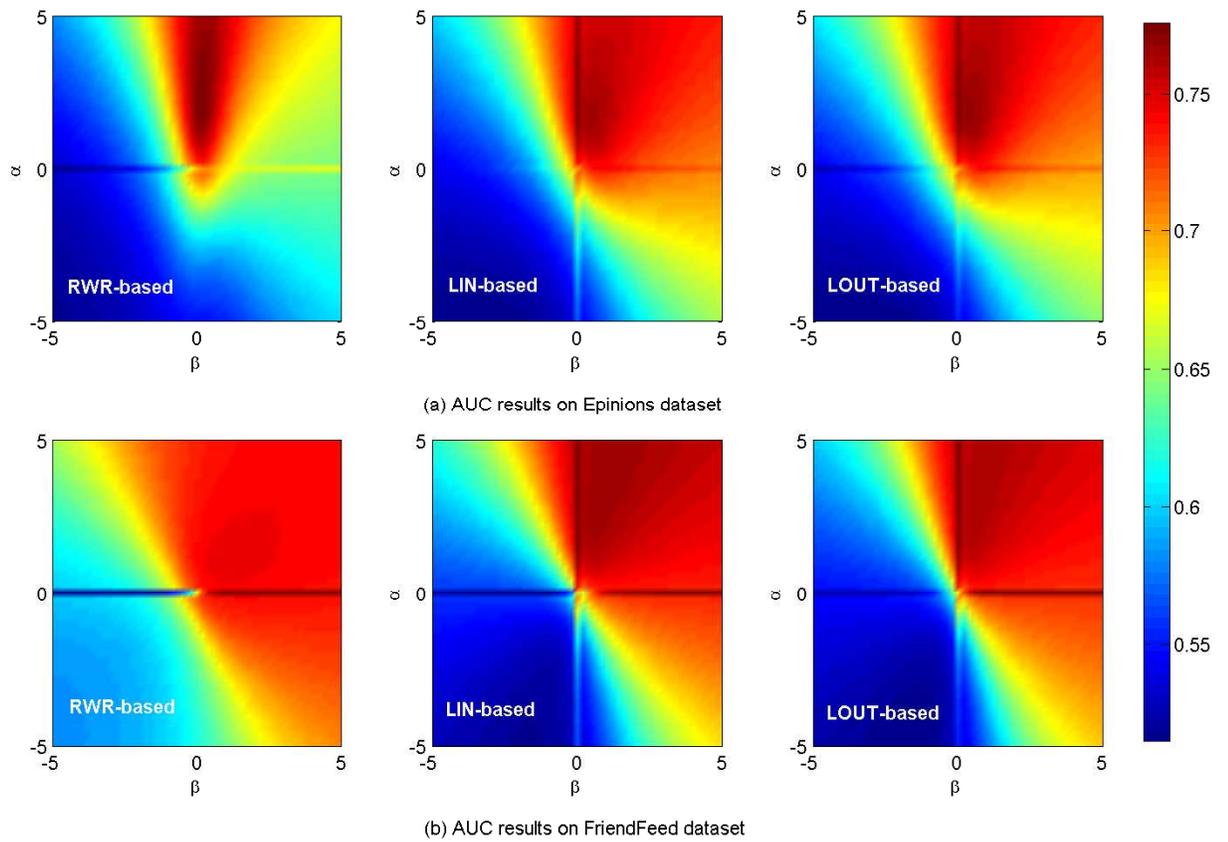}
        \caption{(Color online) \emph{AUC} results on $Epinions$ and $FriendFeed$ data sets.}
    \label{fig_AUC}
    \end{center}
\end{figure}

Fig. \ref{fig_AUC} shows the \emph{AUC} results. In Fig. \ref{fig_AUC}(a), the maximum \emph{AUC} values are respectively 0.7755, 0.7729 and 0.7729 for ($\alpha$, $\beta$) = (2.4, 0.2), ($\alpha$, $\beta$) = (2.2, 0) and ($\alpha$, $\beta$) = (2.2, 0) on $Epinions$ data set. In Fig. \ref{fig_AUC}(b), the corresponding maximum values are respectively 0.9053, 0.8204 and 0.8208 for ($\alpha$, $\beta$) = (0, 2.2), ($\alpha$, $\beta$) = (2.4, 0) and ($\alpha$, $\beta$) = (1.4, 0) on $Friendfeed$, separately. A brief summary is given in Table \ref{tbl:exp}.

It is noticed that, for all aforementioned results 
two crossing lines can be obviously found for LIN- and LOUT-based methods at $\alpha=0$ or $\beta=0$, while only horizonal line is observed for RWR-based method at $\alpha=0$.
As shown in Table \ref{tbl:data}, the information network is much sparser than that of corresponding social network, hence more items are possible to be discovered via social
connections. In addition, the size of hot areas (correspond to high performance) of RWR-based method is much larger than the other two methods, as it considers not only the
nearest neighbors, but also integrates the effect of remote nodes which are not directly connected. Comparatively, the local based (LIN- and LOUT-based)
methods can only take into account the commonly direct neighbors, neglecting the global role of each individual. Furthermore, the hybrid case will reach the best performance for
both the observed data sets with optimal parameters $\alpha^{*} > $ $\beta^{*}$, which also proves that social reinforcement is more significant than individual behaviors in information filtering.


\subsection{Empirical Analysis}
\begin{figure}
\begin{center}
       \center \includegraphics[width=18cm]{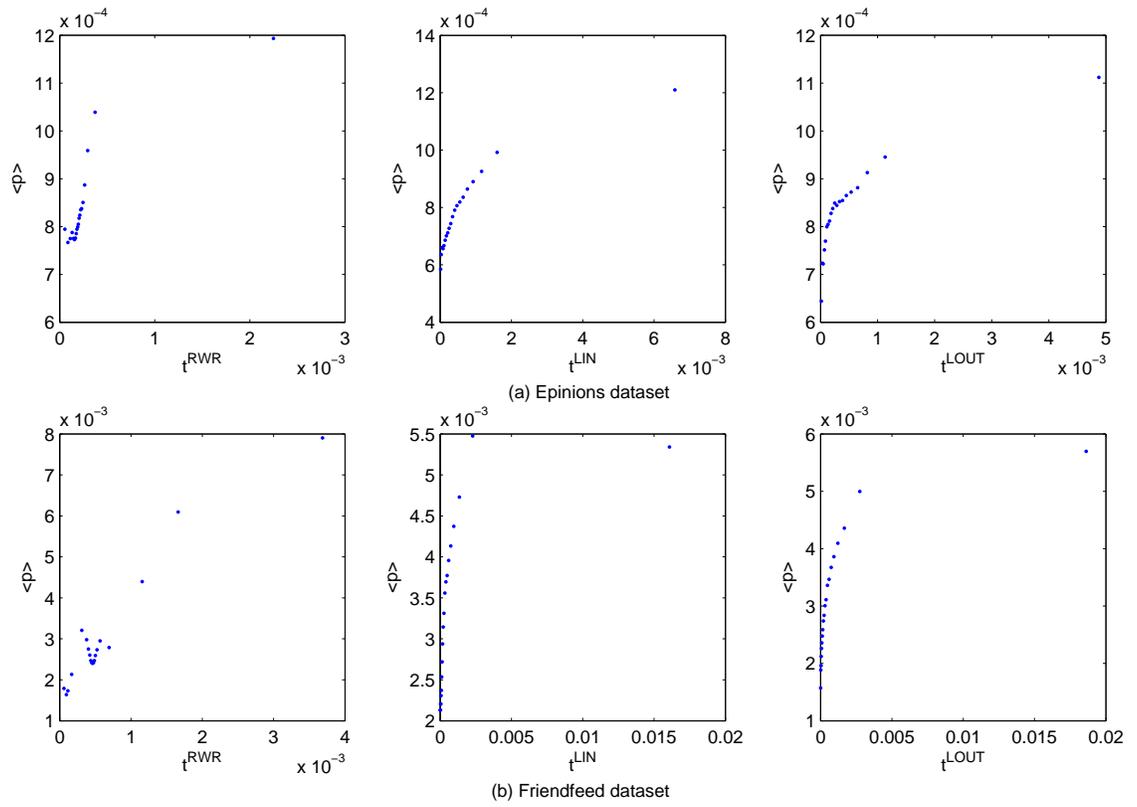}
\caption{Mean personal preference versus social influence for $Epinions$ and $Friendfeed$, respectively. From left to right, the metrics are respectively RWR-, LIN-, LOUT-based social influence. The personal preference is averaged according to each social influence value.} \label{fig:relationship}
\label{fig3}
\end{center}
\end{figure}

To better understand how the different layers of coupled networks interact with each other, in this section, we shall empirically investigate the relationship
between social influence and personal preference from micro/macro perspectives. Fig. \ref{fig:relationship} shows the relationship between social influence and personal preference for each pair of users. It shows that, generally, they are positively correlated \cite{Bhuiyan2010} for both local and global measures, indicating that the mutual reinforcement principle \cite{Kleinberg1999} also applies in online social activities.

\begin{figure}
\begin{center}
       \center \includegraphics[width=16cm]{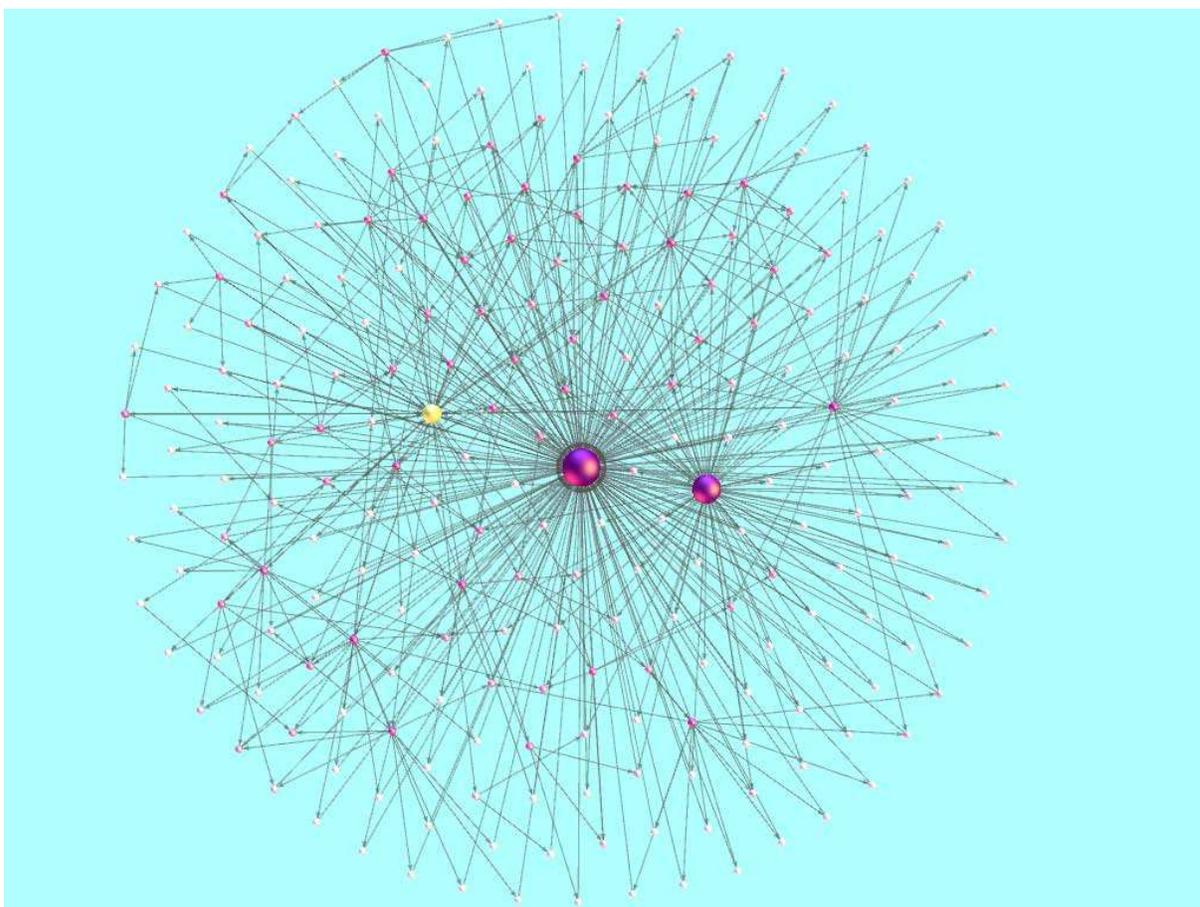}
\caption{(Color online) Illustation of a typical example of an ego network for a node with the largest social influence value (the biggest size).}\label{fig:example}
\label{fig2} \end{center}
\end{figure}

\begin{figure}
\begin{center}
       \center \includegraphics[width=18cm]{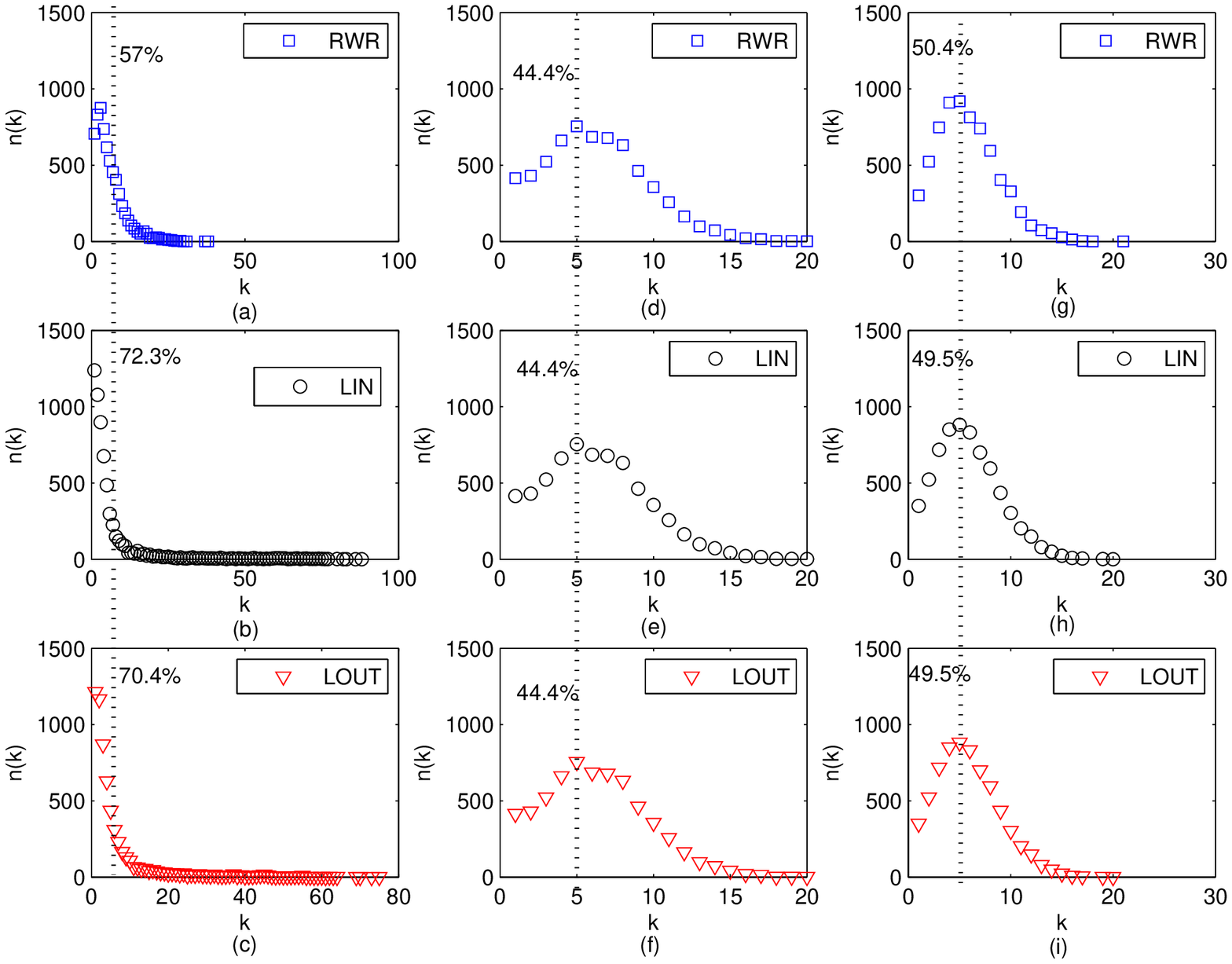}
\caption{Number of recomended items versus degree on \emph{Epinions} for $L=10$. From left to right, the parameters ($\alpha,\beta$) of Eq. (\ref{eq5}) are set as (1,0), (0,1), and ($\alpha^*,\beta^*$) given in Table \ref{tbl:exp}, respectively. The dash line indicates the degree of 5, and the corresponding number shows the its percentage of all the recommenation items.}\label{fig:degree_e}
\end{center}
\end{figure}

\begin{figure}
\begin{center}
       \center \includegraphics[width=18cm]{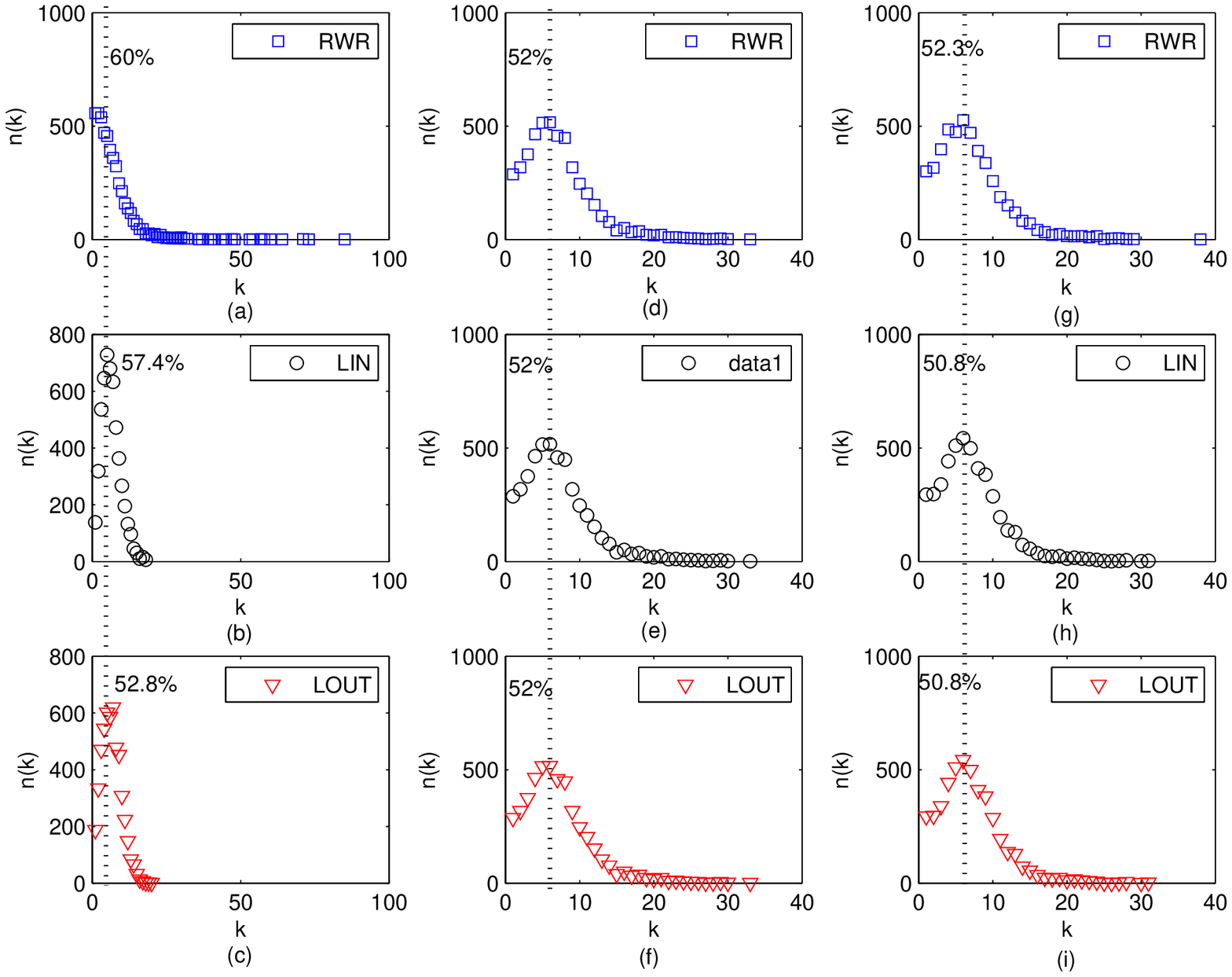}
\caption{Number of recomended items versus degree on \emph{FriendFeed} for $L=10$. From left to right, the parameters ($\alpha,\beta$) of Eq. (\ref{eq5}) are set as (1,0), (0,1), and ($\alpha^*,\beta^*$) given in Table \ref{tbl:exp}, respectively. The dash line indicates the degree of 5, and the corresponding number shows the its percentage of all the recommenation items.}\label{fig:degree_f}
\end{center}
\end{figure}

In Fig. \ref{fig:example}, we also show a typical example of an ego network \cite{Mednick2010} for a node with the largest social influence value (with the biggest size). It can be seen that it connects to a node of relatively large social influence yet small similarity (yellow one), suggesting the rich-club phenomenon \cite{ZhouS2004} of social interests activities. That is to say, users with high social impact tend to interact with users of high social influence, even if they lack common activities. Furthermore, we show the degree distribution of successfully recommended items in Fig. \ref{fig:degree_e} and Fig. \ref{fig:degree_f} for Epinions and Friendfeed, respectively. In Fig. \ref{fig:degree_e}(a-c) and Fig. \ref{fig:degree_f}(a-c), the parameters of Eq. \ref{eq5} are set as $\alpha=0$ and $\beta=1$, of which only the social influence takes effect in the recommendation process. It shows that the local measures (LIN and LOUT) tend to find small-degree items (the degree is smaller than 6) than the RWR metric (around 57\%). Similarity, for another extreme case of Eq. \ref{eq5}, ($\alpha,\beta$) is set as (1,0), implying that only the personal preference will work for information filtering, hence all results are identical in Fig. \ref{fig:degree_e}(d-f) and Fig. \ref{fig:degree_f}(d-f), respectively. In addition, the number of recommended small-degree items are smaller than that of social based method. Comparatively, in Fig. \ref{fig:degree_e}(g-i) and Fig. \ref{fig:degree_f}(g-i), the parameter ($\alpha,\beta$) is set as the optimal case given in Table \ref{tbl:exp}. Since both the social influence and personal preference are integrated, the hybrid algorithm not only can find those \emph{cold} items \cite{qiu2011item, Zhang2010EPL} (where the social influence primarily works), but also can push some popular items (which is largely because of the personal preference). Therefore, it finally can achieve a better performance for information filtering.

\section{Conclusions \& Discussion}
In this paper, we have proposed a hybrid information filtering algorithm based on the coupled social networks, which considers the effects of both social influence and personalized preference. We apply three metrics, \emph{LIN}, \emph{LOUT} and \emph{RWR}, to evaluate the asymmetrically social influence, and use the cosine similarity to measure the symmetrically personalized preference. In addition, we integrate them with two tunable parameters in order to obtain better recommendation results. Experimental results show that hybrid pattern can not only provide more accurate recommendations, but also enlarge the recommendation coverage while adopting global metric (RWR). Further empirical analyses demonstrate that the mutual reinforcement can also be extended to coupled networks where the same individuals occupy the core position of the entire online society. However, This article only provides a simple start for making use of both behavior and social information, while a couple of issues remain open for future study. Especially, the underlying mechanism driving the interaction of social and information networks is of particular importance to deeply understand how couples social networks works, as well as its potential applications.

\section{Acknowledgments}
This work was partially supported by the National Natural Science Foundation of China (Grant Nos. 11105024, 61103109, 1147015, 11301490 and 11305043), the Zhejiang Talents Project (No. QJC1302001), the EU FP7 Grant 611272 (project GROWTHCOM), the start-up foundation and Pandeng project of Hangzhou Normal University.

\section*{Supporting Information Legends}
The data sets are available as attachment: \textbf{Data S1}.


\bibliographystyle{apsrev}
\bibliography{references}

\begin{thebibliography}{10}
\providecommand{\url}[1]{\texttt{#1}}
\providecommand{\urlprefix}{URL }
\expandafter\ifx\csname urlstyle\endcsname\relax
  \providecommand{\doi}[1]{doi:\discretionary{}{}{}#1}\else
  \providecommand{\doi}{doi:\discretionary{}{}{}\begingroup
  \urlstyle{rm}\Url}\fi
\providecommand{\bibAnnoteFile}[1]{%
  \IfFileExists{#1}{\begin{quotation}\noindent\textsc{Key:} #1\\
  \textsc{Annotation:}\ \input{#1}\end{quotation}}{}}
\providecommand{\bibAnnote}[2]{%
  \begin{quotation}\noindent\textsc{Key:} #1\\
  \textsc{Annotation:}\ #2\end{quotation}}
\providecommand{\eprint}[2][]{\url{#2}}

\bibitem{faloutsos1999}
Faloutsos M, Faloutsos P, Faloutsos C (1999) On power-law relationships of the
  internet topology.
\newblock Comput Commun Rev 29: 251--262.
\bibAnnoteFile{faloutsos1999}

\bibitem{froomkin1995}
Froomkin AM (1995) Flood control on the information ocean: Living with
  anonymity, digital cash, and distributed databases.
\newblock Journal of Law and Commerce 15: 395.
\bibAnnoteFile{froomkin1995}

\bibitem{brin1998anatomy}
Brin S, Page L (1998) The anatomy of a large-scale hypertextual web search
  engine.
\newblock Computer networks and ISDN systems 30: 107--117.
\bibAnnoteFile{brin1998anatomy}

\bibitem{morita1994}
Morita M, Shinoda Y (1994) Information filtering based on user behavior
  analysis and best match text retrieval.
\newblock In: Proc. 17th Ann. Intl. ACM SIGIR Conf. Research Develop. Infor.
  Retr. Springer-Verlag New York, Inc., pp. 272--281.
\bibAnnoteFile{morita1994}

\bibitem{lee2008find}
Lee D (2008) To find or to be found, that is the question in mobile information
  retrieval.
\newblock In: Proce. SIGIR 2008 Workshop on Mobile Infor. Retr. pp. 7--10.
\bibAnnoteFile{lee2008find}

\bibitem{resnick1997}
Resnick P, Varian H (1997) Recommender systems.
\newblock Commun ACM 40: 56--58.
\bibAnnoteFile{resnick1997}

\bibitem{burke2002}
Burke R (2002) Hybrid recommender systems: Survey and experiments.
\newblock User modeling and user-adapted interaction 12: 331--370.
\bibAnnoteFile{burke2002}

\bibitem{Herlocker2004}
Herlocker JL, Konstan JA, Terveen LG, Riedl JT (2004) Evaluating collaborative
  filtering recommender systems.
\newblock ACM Transactions on Information Systems 22: 5-53.
\bibAnnoteFile{Herlocker2004}

\bibitem{Adomavicius2005}
Adomavicius G, Tuzhilin A (2005) Toward the next generation of recommender
  systems: A survey of the state-of-the-art and possible extensions.
\newblock IEEE Transactions on Knowledge and Data Engineeing 17: 734-749.
\bibAnnoteFile{Adomavicius2005}

\bibitem{lv2012}
L\"{u} L, Medo M, Yeung CH, Zhang YC, Zhang ZK, et~al. (2012) Recommender
  systems.
\newblock Phys Rep 519: 1--49.
\bibAnnoteFile{lv2012}

\bibitem{huang2004}
Huang Z, Chen H, Zeng D (2004) Applying associative retrieval techniques to
  alleviate the sparsity problem in collaborative filtering.
\newblock ACM Trans Info Syst 22: 116--142.
\bibAnnoteFile{huang2004}

\bibitem{Zhou2007}
Zhou T, Ren J, Medo M, Zhang YC (2007) Bipartite network projection and
  personal recommendation.
\newblock Phys Rev E 76: 0461115.
\bibAnnoteFile{Zhou2007}

\bibitem{breese1998empirical}
Breese J, Heckerman D, Kadie C (1998) Empirical analysis of predictive
  algorithms for collaborative filtering.
\newblock In: Proc. 4th Conf. Uncertainty Artif. Intel. Morgan Kaufmann
  Publishers Inc., pp. 43--52.
\bibAnnoteFile{breese1998empirical}

\bibitem{sarwar2001item}
Sarwar B, Karypis G, Konstan J, Reidl J (2001) Item-based collaborative
  filtering recommendation algorithms.
\newblock In: Proc. 10th Intl. Conf. WWW. ACM, pp. 285--295.
\bibAnnoteFile{sarwar2001item}

\bibitem{pazzani2007content}
Pazzani M, Billsus D (2007) Content-based recommendation systems.
\newblock The adaptive web : 325--341.
\bibAnnoteFile{pazzani2007content}

\bibitem{kumar1998recommendation}
Kumar R, Raghavan P, Rajagopalan S, Tomkins A (1998) Recommendation systems: A
  probabilistic analysis.
\newblock In: Foundations of Computer Science, 1998. Proceedings. 39th Annual
  Symposium on. IEEE, pp. 664--673.
\bibAnnoteFile{kumar1998recommendation}

\bibitem{krestel2009latent}
Krestel R, Fankhauser P, Nejdl W (2009) Latent dirichlet allocation for tag
  recommendation.
\newblock In: Proceedings of the third ACM conference on Recommender systems.
  ACM, pp. 61--68.
\bibAnnoteFile{krestel2009latent}

\bibitem{sarwar2000application}
Sarwar B, Karypis G, Konstan J, Riedl J (2000) Application of dimensionality
  reduction in recommender system-a case study.
\newblock Technical report, DTIC Document.
\bibAnnoteFile{sarwar2000application}

\bibitem{Zhang2007}
Zhang YC, Blattner M, Yu YK (2007) Heat conduction process on community
  networks as a recommendation model.
\newblock Phys Rev Lett 99: 154301-154304.
\bibAnnoteFile{Zhang2007}

\bibitem{burke2002hybrid}
Burke R (2002) Hybrid recommender systems: Survey and experiments.
\newblock User modeling and user-adapted interaction 12: 331--370.
\bibAnnoteFile{burke2002hybrid}

\bibitem{Zhou2010}
Zhou T, Kuscsik Z, Liu JG, Medo M, Wakeling JR, et~al. (2010) Solving the
  apparent diversity-accuracy dilemma of recommender systems.
\newblock Proc Natl Acad Sci USA 107: 18803-18808.
\bibAnnoteFile{Zhou2010}

\bibitem{resnick1994grouplens}
Resnick P, Iacovou N, Suchak M, Bergstrom P, Riedl J (1994) Grouplens: an open
  architecture for collaborative filtering of netnews.
\newblock In: Proc. 1994 ACM Conf. Comput. Supported Cooperative work. ACM, pp.
  175--186.
\bibAnnoteFile{resnick1994grouplens}

\bibitem{linden2003amazon}
Linden G, Smith B, York J (2003) Amazon.com recommendations: Item-to-item
  collaborative filtering.
\newblock IEEE Internet Comput 7: 76--80.
\bibAnnoteFile{linden2003amazon}

\bibitem{zhang2010hypergraph}
Zhang ZK, Liu C (2010) A hypergraph model of social tagging networks.
\newblock J Stat Mech 2010: P10005.
\bibAnnoteFile{zhang2010hypergraph}

\bibitem{Zhang2010EPL}
Zhang ZK, Liu C, Zhang YC, Zhou T (2010) Solving the cold-start problem in
  recommender systems with social tags.
\newblock EPL 92: 28002-28007.
\bibAnnoteFile{Zhang2010EPL}

\bibitem{zhang2011tag}
Zhang ZK, Zhou T, Zhang YC (2011) Tag-aware recommender systems: A
  state-of-the-art survey.
\newblock Journal of Computer Science and Technology 26: 767--777.
\bibAnnoteFile{zhang2011tag}

\bibitem{wang2001efficient}
Wang F, Landau D (2001) Efficient, multiple-range random walk algorithm to
  calculate the density of states.
\newblock Phys Rev Lett 86: 2050--2053.
\bibAnnoteFile{wang2001efficient}

\bibitem{leicht2006vertex}
Leicht E, Holme P, Newman M (2006) Vertex similarity in networks.
\newblock Phys Rev E 73: 026120.
\bibAnnoteFile{leicht2006vertex}

\bibitem{lu2011pre}
L{\"u} L, Liu W (2011) Information filtering via preferential diffusion.
\newblock Physical Review E 83: 066119.
\bibAnnoteFile{lu2011pre}

\bibitem{liujg2011pre}
Liu JG, Zhou T, Guo Q (2011) Information filtering via biased heat conduction.
\newblock Physical Review E 84: 037101.
\bibAnnoteFile{liujg2011pre}

\bibitem{liujg2012pre}
Liu JG, Shi K, Guo Q (2012) Solving the accuracy-diversity dilemma via directed
  random walks.
\newblock Physical Review E 85: 016118.
\bibAnnoteFile{liujg2012pre}

\bibitem{sun2009information}
Sun D, Zhou T, Liu JG, Liu RR, Jia CX, et~al. (2009) Information filtering
  based on transferring similarity.
\newblock Phys Rev E 80: 17101.
\bibAnnoteFile{sun2009information}

\bibitem{qiu2011item}
Qiu T, Chen G, Zhang Z, Zhou T (2011) An item-oriented recommendation algorithm
  on cold-start problem.
\newblock EPL 95: 58003.
\bibAnnoteFile{qiu2011item}

\bibitem{chen2012promotional}
Chen G, Qiu T, Zhang Z (2012) Promotional effect on cold start problem and
  diversity in a data characteristic based recommendation method.
\newblock arXiv:12052822 .
\bibAnnoteFile{chen2012promotional}

\bibitem{Zhou2009}
Zhou T, Su RQ, Liu RR, Jiang LL, Wang BH, et~al. (2009) Accurate and diverse
  recommendations via eliminating redundant correlations.
\newblock New Journal of Physics 11: 123008.
\bibAnnoteFile{Zhou2009}

\bibitem{Zhang2010}
Zhang ZK, Zhou T, Zhang YC (2010) Personalized recommendation via integrated
  diffusion on user-item-tag tripartite graphs.
\newblock Physica A 389: 179-186.
\bibAnnoteFile{Zhang2010}

\bibitem{freeman1979centrality}
Freeman L (1979) Centrality in social networks conceptual clarification.
\newblock Social networks 1: 215--239.
\bibAnnoteFile{freeman1979centrality}

\bibitem{fu2008empirical}
Fu F, Liu L, Wang L (2008) Empirical analysis of online social networks in the
  age of web 2.0.
\newblock Physica A 387: 675--684.
\bibAnnoteFile{fu2008empirical}

\bibitem{nowak2006five}
Nowak MA (2006) Five rules for the evolution of cooperation.
\newblock Science 314: 1560--1563.
\bibAnnoteFile{nowak2006five}

\bibitem{szabo2007evolutionary}
Szab{\'o} G, F{\'a}th G (2007) Evolutionary games on graphs.
\newblock Phy Rep 446: 97--216.
\bibAnnoteFile{szabo2007evolutionary}

\bibitem{fortunato2010community}
Fortunato S (2010) Community detection in graphs.
\newblock Phys Rep 486: 75--174.
\bibAnnoteFile{fortunato2010community}

\bibitem{kamel2007emerging}
Kamel~Boulos M, Wheeler S (2007) The emerging web 2.0 social software: an
  enabling suite of sociable technologies in health and health care education1.
\newblock Health Information \& Libraries Journal 24: 2--23.
\bibAnnoteFile{kamel2007emerging}

\bibitem{kautz1997referral}
Kautz H, Selman B, Shah M (1997) Referral web: combining social networks and
  collaborative filtering.
\newblock Commun ACM 40: 63--65.
\bibAnnoteFile{kautz1997referral}

\bibitem{buldyrev2010catastrophic}
Buldyrev S, Parshani R, Paul G, Stanley H, Havlin S (2010) Catastrophic cascade
  of failures in interdependent networks.
\newblock Nature 464: 1025--1028.
\bibAnnoteFile{buldyrev2010catastrophic}

\bibitem{givoni2006airline}
Givoni M, Banister D (2006) Airline and railway integration.
\newblock Transport Policy 13: 386--397.
\bibAnnoteFile{givoni2006airline}

\bibitem{gao2011networks}
Gao J, Buldyrev S, Stanley H, Havlin S (2011) Networks formed from
  interdependent networks.
\newblock Nature Phys 8: 40--48.
\bibAnnoteFile{gao2011networks}

\bibitem{lu2009similarity}
L{\"u} L, Jin CH, Zhou T (2009) Similarity index based on local paths for link
  prediction of complex networks.
\newblock Phys Rev E 80: 46122.
\bibAnnoteFile{lu2009similarity}

\bibitem{Massa2007}
Massa P, Avesani P (2007) Trust-aware recommender systems.
\newblock In: Proceedings of the 2007 ACM conference on Recommender systems.
  ACM, pp. 17--24.
\bibAnnoteFile{Massa2007}

\bibitem{Guha2004}
Guha R, Kumar R, Raghavan P, Tomkins A (2004) Propagation of trust and
  distrust.
\newblock In: Proceedings of the 13th international conference on World Wide
  Web. ACM, pp. 403--412.
\bibAnnoteFile{Guha2004}

\bibitem{Abdul-Rahman2000}
Abdul-Rahman A, Hailes S (2000) Supporting trust in virtual communities.
\newblock In: System Sciences, 2000. Proceedings of the 33rd Annual Hawaii
  International Conference on. IEEE, p.~9.
\bibAnnoteFile{Abdul-Rahman2000}

\bibitem{Jsang1998}
Knapskog S (1998) A metric for trusted systems.
\newblock In: Proceedings of the 21st National Security Conference. Citeseer,
  pp. 16--29.
\bibAnnoteFile{Jsang1998}

\bibitem{Bhuiyan2010}
Bhuiyan T (2010) A survey on the relationship between trust and interest
  similarity in online social networks.
\newblock Journal of Emerging Technologies in Web Intelligence 2: 291-299.
\bibAnnoteFile{Bhuiyan2010}

\bibitem{Crandall2008}
Crandall D, Cosley D, Huttenlocher D, Kleinberg J, Suri S (2008) Feedback
  effects between similarity and social influence in online communities.
\newblock In: Proceeding of the 14th ACM SIGKDD international conference on
  Knowledge discovery and data mining. ACM, pp. 160--168.
\bibAnnoteFile{Crandall2008}

\bibitem{esslimani2009social}
Esslimani I, Brun A, Boyer A (2009) From social networks to behavioral networks
  in recommender systems.
\newblock In: Intl. Conf. Adv. Social Netw. Anal. Mining (ASONAM'09). IEEE, pp.
  143--148.
\bibAnnoteFile{esslimani2009social}

\bibitem{Tong2006}
Tong H, Faloutsos C, Pan J (2006) Fast random walk with restart and its
  applications.
\newblock In: Proceedings of the Sixth International Conference on Data
  Mining,2006. ICDM '06. Ieee, pp. 613-622.
\bibAnnoteFile{Tong2006}

\bibitem{shang2009relevance}
Shang MS, L{\"u} L, Zeng W, Zhang YC, Zhou T (2009) Relevance is more
  significant than correlation: Information filtering on sparse data.
\newblock EPL (Europhysics Letters) 88: 68008.
\bibAnnoteFile{shang2009relevance}

\bibitem{yu2014reverse}
Yu AW, Mamoulis N, Su H (2014) Reverse top-k search using random walk with
  restart.
\newblock Proceedings of the VLDB Endowment 7.
\bibAnnoteFile{yu2014reverse}

\bibitem{Anderberg1973}
Anderberg MR (1973) Cluster analysis for applications.
\newblock Academic Press .
\bibAnnoteFile{Anderberg1973}

\bibitem{Mild2002}
Mild A, Reutterer T (2002) An improved collaborative filtering approach for
  predicting cross-category purchases based on binary market basket data.
\newblock Journal of Retailing and Consumer Services 10: 123-133.
\bibAnnoteFile{Mild2002}

\bibitem{Kleinberg1999}
Kleinberg JM (1999) Authoritative sources in a hyperlinked environment.
\newblock J ACM 46: 604--632.
\bibAnnoteFile{Kleinberg1999}

\bibitem{ziegler2013recommender}
Ziegler CN (2013) On recommender systems.
\newblock In: Social Web Artifacts for Boosting Recommenders, Springer. pp.
  11--20.
\bibAnnoteFile{ziegler2013recommender}

\bibitem{liu2014discriminant}
Liu C (2014) Discriminant analysis and similarity measure.
\newblock Pattern Recognition 47: 359--367.
\bibAnnoteFile{liu2014discriminant}

\bibitem{Massa2006}
Massa P, Avesani P (2006) Trust-aware bootstrapping of recommender systems.
\newblock In: ECAI 2006 Workshop on Recommender Systems, Riva del Garda, Italy.
  Citeseer, pp. 29--33.
\bibAnnoteFile{Massa2006}

\bibitem{Celli2010}
Celli F, Di~Lascio F, Magnani M, Pacelli B, Rossi L (2010) Social network data
  and practices: The case of friendfeed.
\newblock Advances in Social computing : 346--353.
\bibAnnoteFile{Celli2010}

\bibitem{zeng2013information}
Zeng W, Zeng A, Shang MS, Zhang YC (2013) Information filtering in sparse
  online systems: recommendation via semi-local diffusion.
\newblock PLoS ONE 8: e79354.
\bibAnnoteFile{zeng2013information}

\bibitem{Hanely1982}
Hanley JA, McNeil BJ (1982) The meaning and use of the area under a receiver
  operating characteristic (roc) curve.
\newblock McNeil, Radiology 143: 29-36.
\bibAnnoteFile{Hanely1982}

\bibitem{Mednick2010}
Mednick SC, Christakis NA, Fowler JH (2010) The spread of sleep loss influences
  drug use in adolescent social networks.
\newblock PloS ONE 5: e9775.
\bibAnnoteFile{Mednick2010}

\bibitem{ZhouS2004}
Zhou S, Mondrag{\'o}n RJ (2004) The rich-club phenomenon in the internet
  topology.
\newblock Commun Lett 8: 180--182.
\bibAnnoteFile{ZhouS2004}

\end{thebibliography}

\newpage

\begin{table}
  \centering
  \begin{threeparttable}[b]
\centering \caption{Basic  properties of the two datasets. $|U|$, $|I|$, $N_R$ and $N_S$ respectively represent the number of users, items, ratings and social activeities. $S_r=\frac{R}{|U|\times|I|}$ and $S_p=\frac{S}{|U|\times (|U|-1)}$  denotes the data sparsity of information and social netorks respectively.}\label{tbl:data}
  \begin{tabular}{lrrrrrr}
    \hline
    Data sets   &      $|U|$  &  $|I|$  &     $N_R$    &   $N_S$   &      $S_r$  &      $S_s$  \\
     \hline
     \emph{Epinions}    &     4,066  &  7,649  &       154,122      &      217,071    &     $5.0\times{10^{-3}}$  &     $1.3\times{10^{-2}}$ \\
     \emph{FriendFeed}  &     4,188  &  5,700  &        96,942      &      386,804    &     $4.1\times{10^{-3}}$&     $2.2\times{10^{-2}}$ \\
     \hline
  \end{tabular}
\end{threeparttable}
\end{table}

\begin{table}
\small
  \centering
  \begin{threeparttable}[b]
  \caption{Performance of the recommendation algorithms four metrics: precision (P), recall (R), f-measure (F), and AUC in Epinions and Friendfeed data sets, respectively. $L$ is the length of recommendation list.} \label{tbl:exp}
    \begin{tabular}{|l|c|c|c|c|c|c|c|c|}
    \multicolumn{2}{l}{}&\multicolumn{3}{c}{\textbf{Epinions}} &\multicolumn{1}{c}{}& \multicolumn{3}{c}{\textbf{FriendFeed}} \\
    \cline{1-2}
    \cline{3-5}
    \cline{7-9}
    \multirow{1}{28 pt}{Method} & Metrics & $L=10$ & $L=20$ & $L=50$ & & $L=10$ & $L=20$ & $L=50$\\
    \hline
    \multirow{9}{10 pt}{RWR} &\multirow{2}{10 pt}{P}& \textbf{0.0526} & \textbf{0.0402} & \textbf{0.0273} & & \textbf{0.0425} & \textbf{0.0325} & \textbf{0.0231}\\
    & & (2.8, 0.4) & (2.6, 0.4) & (2.8, 0.2) & & (2, 0.8) & (1.8, 1.2) & (1.6, 1)\\
    \cline{2-9}
    &\multirow{2}{10 pt}{R}& \textbf{0.0717} & \textbf{0.1076} & \textbf{0.1776} & & \textbf{0.1006} & \textbf{0.1507} & \textbf{0.2550} \\
    & & (2.8, 0.4)  & (2.2, 0.4)  & (2.4, 0.2)  & & (1.4, 0.8) & (1.4, 0.4)  & (1.6, 1) \\
    \cline{2-9}
    &\multirow{2}{10 pt}{F}& \textbf{0.0512} & \textbf{0.0503} & \textbf{0.0426} & & \textbf{0.0469} & \textbf{0.0435} & \textbf{0.0370}\\
    & & (2.8, 0.4)  & (2.6, 0.4)  & (2.4, 0.2)  & & (2, 0.8) & (1.6, 1)  & (1.6, 1) \\
    \cline{2-9}
    &\multirow{1}{10 pt}{AUC}& \multicolumn{3}{c|}{\textbf{0.7755} (2.4, 0.2)} & & \multicolumn{3}{c|}{\textbf{0.9053} (0, 2.2)}\\
    \hline
    \multirow{9}{10 pt}{LIN} &\multirow{2}{10 pt}{P}& 0.0503 & 0.0393 & 0.0270 & & 0.0403 & 0.0311 & 0.0221 \\
    & & (3, 0) & (3.2, 0)  & (2.8, 0)  & & (2.4, 0)  & (2.4, 0)   & (2, 0) \\
    \cline{2-9}
    &\multirow{2}{10 pt}{R}& 0.0683 & 0.1043 & 0.1736 & & 0.0963 & 0.1441 & 0.2399\\
    & & (3, 0)  & (2.6, 0)  & (2.8, 0)  & & (2.2, 0)  & (1.8, 0)  & (2, 0) \\
    \cline{2-9}
    &\multirow{2}{10 pt}{F}& 0.0489 & 0.0487 & 0.0421 & & 0.0443 &0.0414 & 0.0352 \\
    & & (3, 0) & (3.2, 0) & (2.8, 0)  & &  (2.4, 0) &  (2, 0) & (2, 0) \\
    \cline{2-9}
    &\multirow{1}{10 pt}{AUC}& \multicolumn{3}{c|}{0.7729 (2.2, 0)} & & \multicolumn{3}{c|}{0.8204 (2.4, 0)}\\
    \hline
    \multirow{9}{10 pt}{LOUT} &\multirow{2}{10 pt}{P}& 0.0503 & 0.0393 & 0.0270 & & 0.0403 & 0.0311 & 0.0221 \\
    & & (3, 0) & (3.2, 0)  & (2.8, 0)  & & (2.4, 0)  & (2.4, 0)  & (2, 0) \\
    \cline{2-9}
    &\multirow{2}{10 pt}{R}& 0.0683 & 0.1043 & 0.1736 & & 0.0963 & 0.1441 & 0.2399 \\
    & & (3, 0) & (2.6, 0) & (2.8, 0)  & & (2.2, 0)  & (1.8, 0)  & (2, 0) \\
    \cline{2-9}
    &\multirow{2}{10 pt}{F}& 0.0489 & 0.0487 & 0.0421 & & 0.0443 & 0.0414 & 0.0352\\
    & & (3, 0)  & (3.2, 0) & (2.8, 0) & & (2.4, 0)  & (2, 0)  & (2, 0) \\
    \cline{2-9}
    &\multirow{1}{10 pt}{AUC}& \multicolumn{3}{c|}{0.7729 (2.2, 0)} & & \multicolumn{3}{c|}{0.8208 (1.4, 0)}\\
    \hline
    \end{tabular}
\end{threeparttable}
\end{table}

\clearpage

\section*{Figure Legends}

\newpage

\newpage

\newpage

\newpage

\newpage

\newpage

\newpage

\newpage

\newpage

\end{document}